\documentclass[aps,prd,twocolumn,tightenlines,nofootinbib,superscriptaddress,floatfix]{revtex4-2}
 
\usepackage{subcaption}
\usepackage{slashed}
\usepackage{hyperref}  
\usepackage{amsmath}
\usepackage{feynmp-auto} 
\usepackage{graphicx} 

\begin{document}

\title{New Physics Searches via Beam Normal Spin Asymmetry in Bhabha Scattering}

\author{Aleksandr Pustyntsev}
\affiliation{Institut f\"ur Kernphysik and $\text{PRISMA}^{++}$ Cluster of Excellence, Johannes Gutenberg Universit\"at, D-55099 Mainz, Germany}
\author{Muthubharathi S. Ramasamy}
\affiliation{Institut f\"ur Kernphysik and $\text{PRISMA}^{++}$ Cluster of Excellence, Johannes Gutenberg Universit\"at, D-55099 Mainz, Germany}
\author{Marc Vanderhaeghen}
\affiliation{Institut f\"ur Kernphysik and $\text{PRISMA}^{++}$ Cluster of Excellence, Johannes Gutenberg Universit\"at, D-55099 Mainz, Germany}

\date{\today}

\begin{abstract}
We examine the sensitivity of the beam normal spin asymmetry in Bhabha scattering to beyond the Standard Model (BSM) mediators, in the context of the JLab polarized positron program. A key property of this observable is that the Standard Model contribution exhibits a zero crossing at a fixed scattering angle, providing a clean, effectively background-free point for these searches. We consider scalar, vector, and axial vector mediators and present projected bounds, finding that scalar and vector scenarios allow a significant extension of the search ranges beyond existing constraints.
\end{abstract}

\maketitle

\section{Introduction}\label{sec1}

The search for light dark sector particles in the MeV to GeV mass range has intensified over the past decade \cite{Beacham:2019nyx,Lanfranchi:2020crw,Agrawal:2021dbo,Batell:2022dpx,Antel:2023hkf,Balan:2024cmq}, with a particular focus on axion-like particles (ALPs) \cite{Jaeckel:2015jla,Izaguirre:2016dfi,Knapen:2016moh,Bauer:2017ris,Dolan:2017osp,Merlo:2019anv,NA64:2020qwq,ATLAS:2020hii,Belle-II:2020jti,BaBar:2021ich,BESIII:2022rzz,Pustyntsev:2023rns,RebelloTeles:2023uig,BESIII:2024hdv,Pustyntsev:2024ygw} and dark photons \cite{APEX:2011dww,Merkel:2014avp,NA482:2015wmo,KLOE-2:2016ydq,Chang:2016ntp,Feng:2017drg,HPS:2018xkw,LHCb:2019vmc,Adrian:2022nkt,BESIII:2022oww,NA62:2023qyn,FASER:2023tle,NA64:2023wbi}. 
The coverage across this mass range is primarily provided by $e^+e^-$ colliders, beam-dump, and fixed-target experiments, mostly relying on “bump hunt” and missing energy searches. These have proven powerful but are not exhaustive. In particular, the few MeV to few-hundred MeV mediator-mass range remains relatively underexplored \cite{Belle-II:2020jti,BESIII:2022rzz,BESIII:2024hdv}. The detection of low-energy final states is challenged by large QED backgrounds and detector-threshold effects which complicate reconstruction \cite{Dolan:2017osp,Pustyntsev:2023rns,Pustyntsev:2024ygw}, motivating complementary low-energy precision programs such as PADME~\cite{PADME:2025dla}, the MAGIX@MESA facility \cite{Schlimme:2024eky}, and the Jefferson Lab (JLab) polarized positron program \cite{Accardi:2020swt}.

The latter is the main subject of this work. While existing studies have placed constraints on BSM effects using unpolarized Bhabha scattering \cite{Bourilkov:1999iz,Bourilkov:2000ap,L3:2000bql}, the uniqueness of the JLab positron project stems from its high-intensity positron beam with a high degree of polarization. The interaction of this beam with unpolarized atomic electrons in a fixed target enables access to polarization-sensitive observables in Bhabha scattering $e^+ e^- \to e^+ e^-$. The beam normal spin asymmetry, which is proportional to an absorptive (imaginary) part of the scattering amplitude, is of particular interest. 

The 11 GeV beam at JLab will allow access to the Bhabha process up to center-of-momentum (c.m.) energies $\sqrt{s} \simeq 106$~MeV. Being below any hadron production threshold this process is theoretically very clean as it cannot get contributions from any hadronic absorptive part. Moreover, the leading QED contribution to this asymmetry vanishes at a fixed scattering angle, yielding a kinematic point with strongly reduced SM background for BSM searches. Since QED interactions at energies much larger than the electron mass are helicity-conserving, particularly strong bounds can be derived for dark sector mediators which induce a lepton helicity flip, such as scalars. Furthermore, our findings show that the projected dark photon search range can also be extended significantly beyond the existing limits, demonstrating the complementarity and broad potential of $e^+e^-$ polarization observables in light messenger particle searches.

The paper is organized as follows. In Section \ref{sec2}, we outline the beam normal spin asymmetry in Bhabha scattering. Section \ref{sec3} details the QED background to this observable, while in Section \ref{sec4} we investigate the BSM contributions for generic scalar, vector, and axial vector mediators. Section \ref{sec5} presents the sensitivity projections for JLab kinematics. Finally, section \ref{sec6} provides a summary of our work.

\section{Beam normal spin asymmetry $B_n$}\label{sec2}

In this work, we consider the Bhabha scattering process
\begin{equation}
e^-\left(p_1, s_1\right) + e^+\left(p_2, s_2\right) \to e^-\left(p_3, s_3\right) + e^+\left(p_4, s_4\right),
\end{equation}
where $p_1$ and $p_2$ ($p_3$ and $p_4$) denote the initial (final) electron and positron four-momenta, and the $s_i$ denote the corresponding helicities. 

\begin{figure*}
    \centering
    \includegraphics[width=0.825\linewidth]{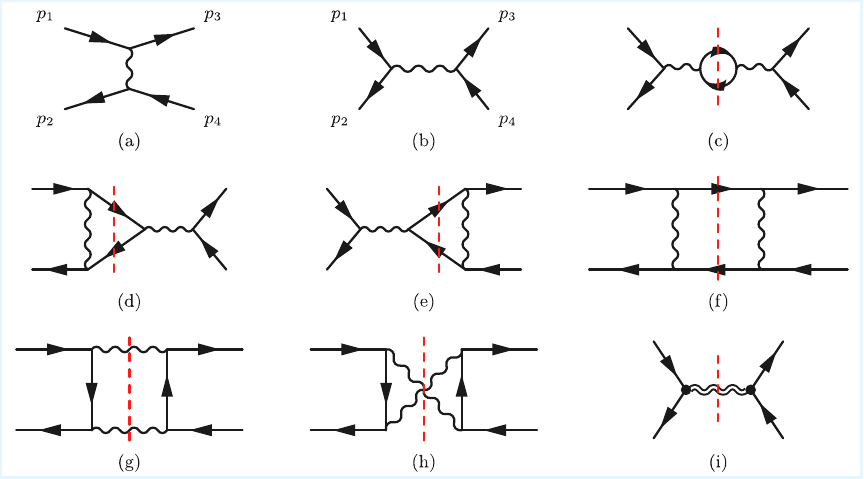}
    \caption{Feynman diagrams contributing to $B_n$ at one-loop order (a-h), and the BSM contribution (i).  
The red vertical line shows the physical-region cuts.}
    \label{fig:FD}
\end{figure*}

Imposing parity (P), time-reversal (T), and charge-conjugation (C) invariance leads to five independent helicity amplitudes for this process~\cite{GOLDBERGER1957226}. 
For the purpose of exploring the contribution of P, T, and C conserving dark messenger particles to this process it is convenient to parameterize the amplitude $\mathcal{M}$ through the following covariants:
\begin{equation}\label{eq:expand}
\begin{split}
& \mathcal{M} = S \, \bar{u}_3u_1 \, \bar{v}_2 v_4+ P \,  \bar{u}_3\gamma^5u_1 \, \bar{v}_2\gamma^5 v_4\\
& + V \,  \bar{u}_3 \gamma_{\mu}u_1 \,  \bar{v}_2 \gamma^{\mu}v_4 + A\,  \bar{u}_3 \gamma_{\mu}\gamma^5u_1 \, \bar{v}_2 \gamma^{\mu}\gamma^5 v_4\\
& + T \,  \bar{u}_3 \sigma_{\mu\nu}u_1 \, \bar{v}_2 \sigma^{\mu\nu}v_4,    \\
\end{split}
\end{equation}
where for compactness we adopt the notation  
$u_i \equiv u(p_i,s_i)$ and $v_i \equiv v(p_i,s_i)$ for the Dirac spinors, and use $\sigma^{\mu \nu} = i/2 \, [ \gamma^\mu, \gamma^\nu]_-$ in terms of the Dirac matrices. In Eq.~\eqref{eq:expand}, the invariant amplitudes $S$, $P$, $V$, $A$, and $T$—corresponding to the scalar, pseudoscalar, vector, axial vector, and tensor structures, respectively—are functions of the Mandelstam variables, defined in the standard way
\begin{equation}
s = \left(p_1+p_2\right)^2, \quad t = \left(p_1-p_3\right)^2, \quad u = \left(p_1-p_4\right)^2. 
\end{equation}

The parameterization~\eqref{eq:expand} is arguably the most natural choice—particularly in the context of BSM searches—although it is not unique. We also note that while the decomposition above is written in terms of scattering-channel structures\footnote[1]{The $t$-channel is chosen since, in the kinematic regime relevant to the JLab measurement, its contribution dominates across most scattering angles.} $\bar u_3\Gamma_i u_1\,\bar v_2\Gamma_i v_4$, the annihilation channel structures $\bar v_2\Gamma_i u_1\,\bar u_3\Gamma_i v_4$ can be straightforwardly projected onto these using Fierz identities \cite{Peskin:1995ev}.

The single-spin asymmetry is defined through the beam-polarized cross sections $d\sigma_\uparrow$ and $d\sigma_\downarrow$, differential in the c.m. scattering angle $\theta$, as
\begin{equation}
B_n = \frac{d\sigma_{\uparrow} - d\sigma_{\downarrow}}{d\sigma_{\uparrow} + d\sigma_{\downarrow}}.
\end{equation}
It is expressed in terms of the invariants of Eq. \eqref{eq:expand} as
\begin{equation}
B_n = \left(\mathcal{S} \cdot \hat n \right) \,\frac{m_e }{4\pi s}\frac{\sqrt{stu}}{\sigma_0} \, \text{Im} \left[\left(S-A\right)\left(V^* + 2T^*\right)\right],
\label{eq:Bn}
\end{equation}
with $m_e$ the lepton mass, $\sigma_0$ the unpolarized cross section, $\mathcal{S}^\mu$ the beam particle polarization four-vector, and 
\begin{equation}
\hat n_\mu \equiv \frac{2}{\sqrt{s t u}} \; \varepsilon_{\mu \alpha \beta \gamma} p_1^\alpha p_2^\beta p_3^\gamma,
\end{equation}
where we adopt the choice $\varepsilon_{0123} = +1$ for the antisymmetric Levi-Civita tensor.
In the c.m. frame the four-vector $n^\mu$ can be expressed as $\hat n^\mu = (0, \mathbf{ \hat n})$, with $\mathbf{\hat n}$ the unit three-vector normal to the scattering plane. One thus sees that the factor $\left( \mathcal{S} \cdot \hat n\right)$ in Eq. \eqref{eq:Bn} leads to a non-zero single spin asymmetry for a polarization normal to the scattering plane. 
In the following the asymmetry is shown using the sign convention 
\begin{equation}
\mathcal{S} \cdot \hat n = +1. 
\end{equation}
The normal beam spin asymmetry Eq.~\eqref{eq:Bn} requires the absorptive (imaginary) part of the amplitude to be non-zero. Furthermore, one notices that an amplitude containing a pseudoscalar exchange—the $P$ term in Eq.~\eqref{eq:expand}—does not contribute to $B_n$, as all resulting Dirac traces involving $\gamma^5$ identically vanish in the cross section difference $d\sigma_{\uparrow} - d\sigma_{\downarrow}$, given the three unpolarized leptons and the spin projection operator.  

\section{QED contributions to $B_n$}\label{sec3}

We now discuss subsequently the QED contribution to $B_n$ as well as the $s$-channel resonance excitation of a BSM dark sector messenger particle of different quantum numbers. The tree-level QED diagrams to the Bhabha process are shown in Figs. \hyperref[fig:FD]{1a} and \hyperref[fig:FD]{1b}. Projected on the basis of Eq.~\eqref{eq:expand} they yield the amplitudes:
\begin{eqnarray}
V &=& - e^2\left( \frac{1}{t} + \frac{1}{2s} \right), \nonumber \\
S &=& - P = -2 A = \frac{e^2}{s}, \quad \quad T = 0,
\label{eq:QEDtree}
\end{eqnarray}
where $e$ is the electric charge. 
As the tree-level QED diagrams do not contribute to the imaginary part of the Bhabha scattering amplitude, $B_n$ is zero at tree-level in QED. The leading nonvanishing contribution to the observable arises from the term $\propto 2\text{Im}[\mathcal{M}_{\text{tree}}\mathcal{M}_{\text{1-loop}}^*]$, involving the tree-level amplitude and 1-loop corrections. The relevant 1-loop diagrams generating these amplitudes are

\begin{itemize}
  \item $s$-channel vacuum polarization with intermediate $e^+e^-$ loop, Fig. \hyperref[fig:FD]{1c},
  \item $s$-channel vertex corrections on both the initial- and final-state fermion lines combined with the $t$-channel two-photon box diagrams, for which only the direct graph yields a nonzero imaginary part, Figs. \hyperref[fig:FD]{1d}-\hyperref[fig:FD]{f},
  \item $s$-channel direct and crossed two-photon box diagrams, Figs. \hyperref[fig:FD]{1g} and \hyperref[fig:FD]{1h}.
\end{itemize}
All other 1-loop diagrams have vanishing discontinuities and therefore do not contribute to $B_n$. 
A number of cancellations were found among the individual contributions. The vertex correction diagrams and $t$-channel box diagrams of Figs. \hyperref[fig:FD]{1d}-\hyperref[fig:FD]{f} require an infrared regulator, which cancels out when calculating their contribution to $B_n$, rendering it an infrared-safe observable. Furthermore, any logarithmic dependence on the electron mass cancels out in the final result. 

We calculated the relevant 1-loop QED diagrams and projected them onto the Dirac basis in Eq. \eqref{eq:expand}. Our result for $B_n$ obtained from Eq.~\eqref{eq:Bn} verifies the expression given in the literature~\cite{PhysRev.121.916}.
In the ultra-relativistic limit the resulting QED expression at 1-loop level can be written in the compact form:
\begin{equation}\label{eq:QED}
\begin{split}
& B_n^{\text{QED, 1-loop}} = \alpha\, \frac{m_e}{\sqrt{s}} \,\frac{x}{2\sqrt{1-x^2}\left(1-x^2+x^4\right)^2} \\
& \times \biggl[2 x^2 \left(1-x^2\right) \left(-5 + x^4\right) -3 x^2 \left(1 + x^2\right)\, \ln{\left(x^2\right)}   \\
& + 3 \left(1 - x^2\right)\left(1 - 2x^2\right)\, \ln{ \left(1-x^2\right)}  + \mathcal{O}\left(\frac{m_e^2}{s}\right) \biggr],
\end{split}
\end{equation}
with $\alpha \equiv e^2 /(4 \pi) \simeq 1/137$,  
and $x\equiv\sin{\left(\theta/2\right)}$.

\begin{figure}
    \centering
    \includegraphics[width=0.95\linewidth]{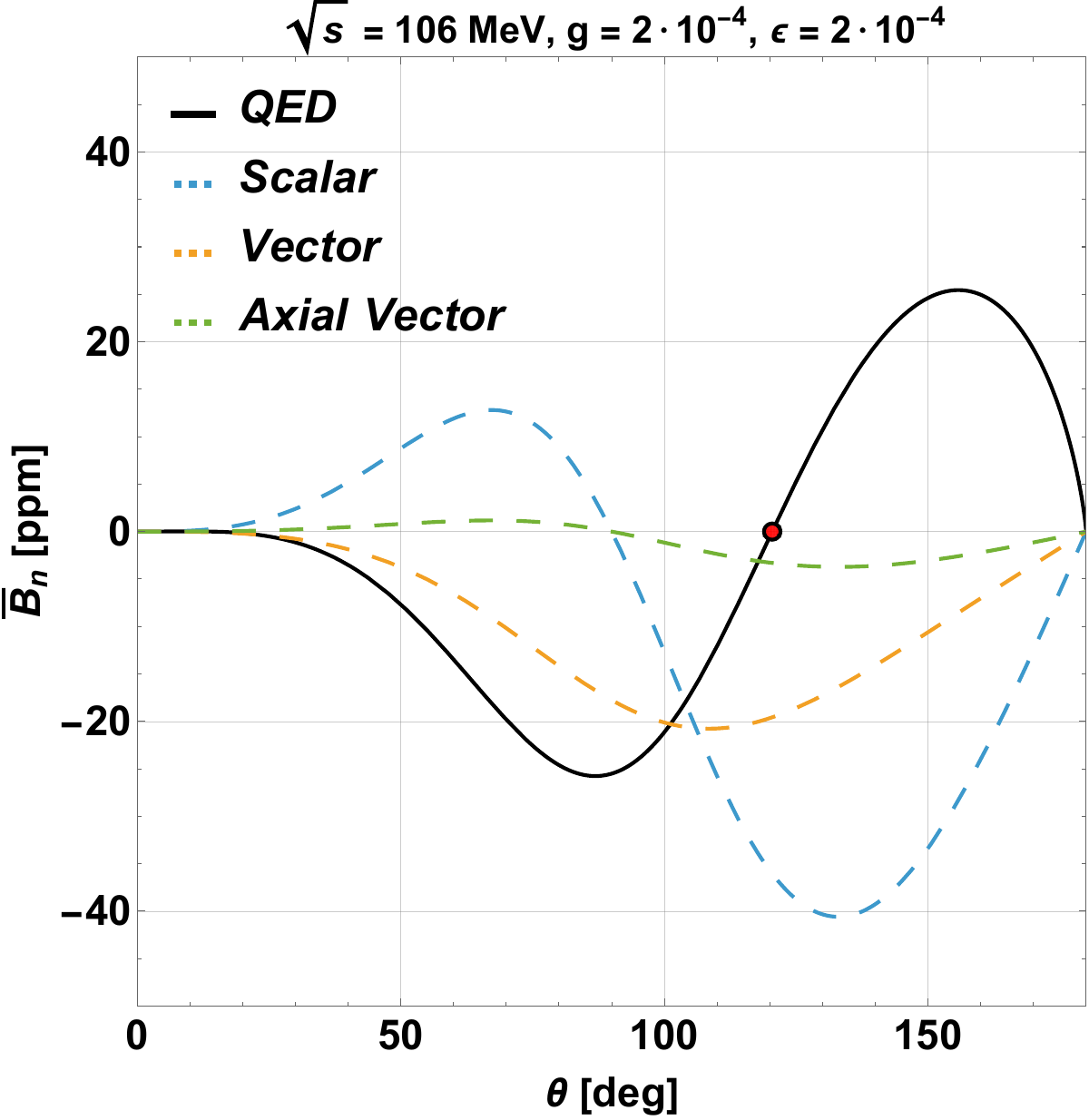}
    \caption{Comparison of the angular dependence of the QED and BSM contributions to the Bhabha beam normal spin asymmetry $\bar{B}_n$ for indicated values of BSM coupling strengths, for an 11 GeV $e^+$ beam at JLab. 
    The numerator and denominator in $\bar{B}_n$ are integrated over a bin in $s$ of size $2 m_e \delta E_+$, with $\delta E_+ = 0.5$~MeV.   
    The QED zero-crossing point is marked with a red dot.}
    \label{fig:BnPlot}
\end{figure}

The QED result for $B_n$, shown on Fig.~\ref{fig:BnPlot},  exhibits a zero crossing at $\theta \approx 120.4^{\circ}$ due to cancellations between different discontinuities. In the ultra-relativistic regime, where the only kinematic dependence appears in the overall prefactor, the position of this zero is independent of $\sqrt{s}$, thus providing a kinematic point that is effectively background free. 

While two-loop effects or finite-mass corrections could in principle shift this zero, these contributions are expected to lie well below the anticipated experimental precision for $B_n$, estimated at $\sim 1$ ppm~\cite{Mack:2025}. Furthermore, a real experiment will measure over a finite angular bin $\Delta\theta$, expected to be $10^\circ$ or smaller. Integrating the smoothly varying SM distribution over a symmetric bin centered on the zero-crossing cancels the SM background to first order. Other potential backgrounds from positron-nucleus elastic scattering $e^+N \to e^+N$ can be effectively eliminated by utilizing the distinct kinematics of the Bhabha process, specifically by requiring a coincidence detection of the final state $e^+e^-$ pair.

\begin{figure*}
    \centering
    \includegraphics[width=0.43\linewidth]{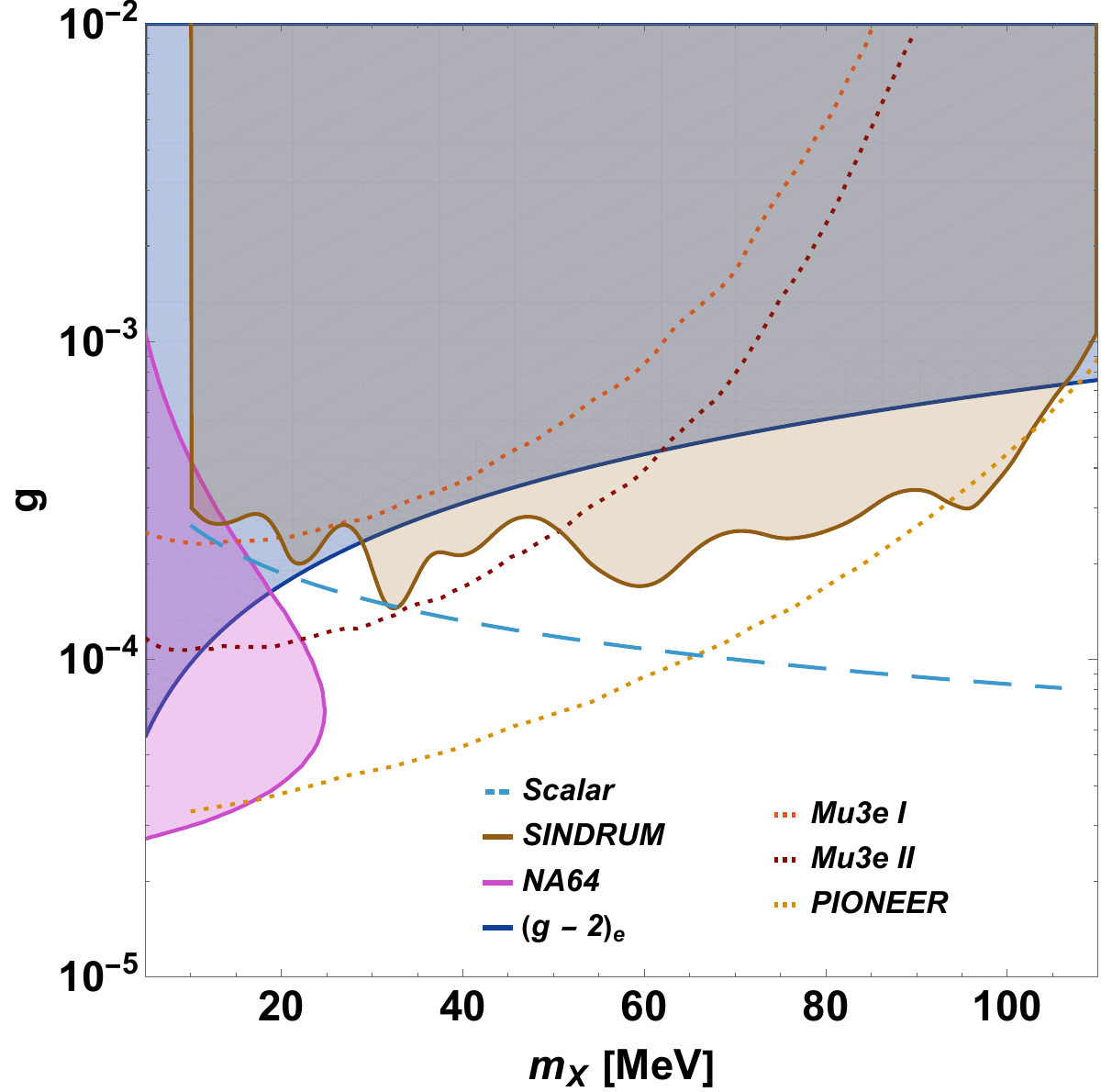} \quad \quad \quad
    \includegraphics[width=0.43\linewidth]{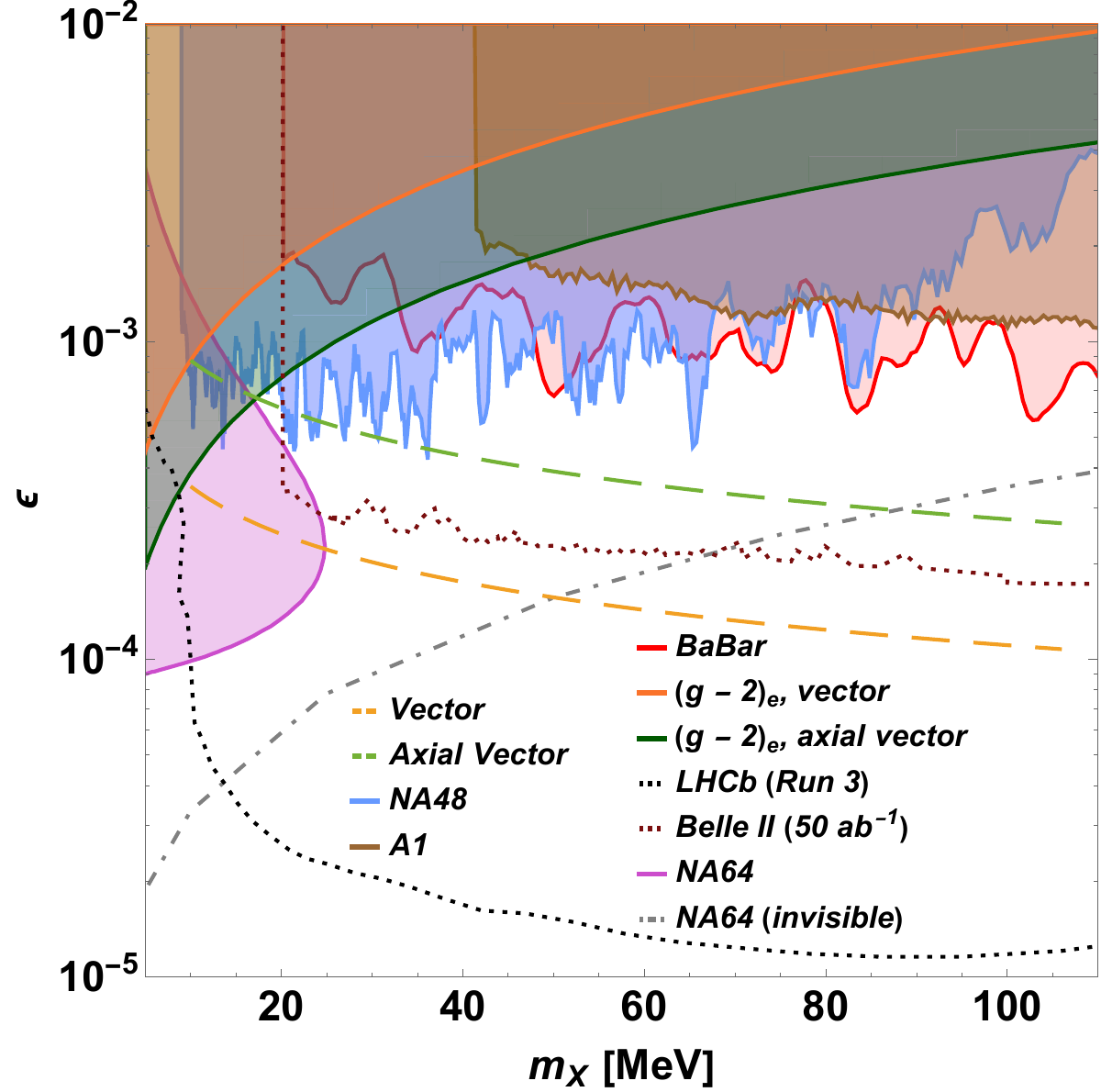}
    \caption{Constraints on the allowed BSM parameter space from $\left(g-2\right)_e$ and from existing experiments searching for visible $e^+e^-$ decays (shaded exclusion regions), shown for the scalar (left) and the vector/axial vector scenarios (right). The projected sensitivity of a future JLab polarized-positron experiment, assuming a 1 ppm precision on $B_n$, is indicated by dashed lines, while other near-future experiments are shown as dotted lines and are described in the text. For comparison, the gray dot-dashed curve shows limits from NA64 searches for invisibly decaying mediators.}
    \label{fig:ExLimit}
\end{figure*}

\section{BSM contributions to $B_n$}\label{sec4}

We next discuss the sensitivity of $B_n$ to BSM messengers. As the exchange of a pseudoscalar particle provides no contribution to $B_n$, seen from Eq.~\eqref{eq:Bn}, three BSM mediators with minimal couplings to SM leptons are considered below: a scalar, a vector, and an axial vector. Their interaction terms are
\begin{equation}
\begin{split}
\mathcal{L}_{S} &= -\, g\, \tilde{S}\, \bar{\ell}\ell ,\\[2mm]
\mathcal{L}_{V} &= -\, e\,\epsilon\, \bar{\ell}\,\slashed{V}\, \ell ,\\[2mm]
\mathcal{L}_{A} &= -\, e\,\epsilon\, \bar{\ell}\,\gamma^{5}\slashed{A}\, \ell ,
\end{split}
\end{equation}
where $\tilde{S}$ is a scalar field coupled to SM leptons with strength $g$, and $V^\mu$ ($A^\mu$) denote the vector (axial vector) mediator fields respectively, with $\epsilon$ due to  kinetic mixing. 
The considered BSM contribution to $B_n$ arises from the term $\propto 2\text{Im}[\mathcal{M}_{\text{tree}}\mathcal{M}_{\text{BSM}}^*]$, involving the tree-level QED amplitude (Figs. \hyperref[fig:FD]{1a} and \hyperref[fig:FD]{1b}) and the $s$-channel tree-level BSM amplitude (Fig. \hyperref[fig:FD]{1i}), with the imaginary part resulting from the propagator
\begin{equation}
\frac{1}{s - m_X^{2} + i m_X \Gamma_X},
\end{equation}
where $m_X$ and $\Gamma_X$ denote the mediator mass and total decay width. In the kinematic region of interest, the only accessible final states are $e^+e^-$ and $\gamma\gamma$. While the Landau-Yang theorem forbids massive spin-1 particles from decaying into two photons, this channel remains open for scalars. However, the corresponding decay width scales as $m_S^3$, in contrast to the leptonic channel, which scales linearly with $m_S$, as discussed in \cite{Pustyntsev:2024ygw}. In the relatively low-mass range studied here, the diphoton mode is therefore strongly suppressed. Consequently, in the minimal model considered here the total width consists of a single contribution from the mediator's coupling to electrons.

Existing constraints imply that $\Gamma_X$ is a small quantity—well below the experimental resolution of the invariant mass of the decay products—so the narrow width approximation for the imaginary part amounts to the replacement
\begin{equation}
\frac{- i m_X \Gamma_X}{(s - m_X^{2})^{2} + m_X^{2}\Gamma_X^{2}} \to
-i \pi \,\delta(s - m_X^{2}).
\end{equation}

To determine the sensitivity reach of the dark messenger coupling as a function of its mass $m_X$, the differential cross sections in the numerator and denominator of $B_n$ must be integrated over a bin in the $e^+e^-$ invariant mass around a given central value $s_0$. For a smoothly varying function, such as the QED cross section, this amounts to the factor
\begin{equation}
\int_{s_0 - \frac{\delta s}{2}}^{s_0 + \frac{\delta s}{2}} f\left(s\right) \text{d}s = 2m_e \,\delta E_+ \, f\left(s_0\right),
\end{equation}
with $\delta E_+$ the bin width in the positron ({\it lab})  beam energy. 
To remain conservative, we adopt a bin width of $\delta E_+ = 0.5 \, \text{MeV}$. For the QED contribution, this factor cancels out between  numerator and denominator, but it becomes relevant for the BSM-induced result. Denoting  the asymmetry in which numerator and denominator are integrated over a bin in $s$ of size $2 m_e\,\delta E_+$ as $\bar{B}_n$, we find that this results in
\begin{eqnarray}
\bar{B}_n^{\text{BSM}} 
= \frac{\sqrt{s}}{\delta E_+} \frac{x\sqrt{1-x^2}}{8} \frac{N_i}{\left(1-x^2+x^4\right)^2},
\end{eqnarray}
where the numerator function $N_i$ for scalar ($N_S$), vector ($N_V$), and axial vector ($N_A$) exchange is given by:
\begin{align}
& N_S = \frac{g^2}{\alpha} \, x^2 \left(1 - 2 x^2\right), 
\label{eq:NS}\\
& N_V = - 4\pi \epsilon^2 \, 3x^2, 
\label{eq:NV}\\
& N_A = 4\pi\epsilon^2 \, x^2 \left(1 - 2 x^2\right).
\label{eq:NA}
\end{align}

Note that in the ultra-relativistic regime the scalar and axial vector contributions have the same angular ($x$) dependence up to the coupling constant. In contrast to the QED contribution, which decreases with energy as $1/\sqrt{s}$, the BSM-induced spin asymmetry grows proportionally to $\sqrt{s}$, improving the signal-to-background ratio at higher beam energies. Furthermore, one notices from Eqs.~\eqref{eq:NS}-\eqref{eq:NA} that the observable $B_n$ depends quadratically on the BSM couplings, as it results from the product of a large tree-level QED amplitude and the absorptive part of a BSM amplitude. This is in contrast to BSM searches in unpolarized observables, for which the BSM signal goes with the fourth power of the small coupling constant. One thus sees that the tree-level QED process serves as an amplifier of a possible BSM signal.  

The result for the angular dependence of $B_n$ is shown in Fig. \ref{fig:BnPlot} for an 11 GeV positron beam at JLab. The coupling strengths $g = 2\times10^{-4}$ and $\epsilon = 2\times10^{-4}$ are chosen for illustration. Even for such small values, one sees that the BSM effects in the vicinity of the QED zero crossing can be sizable compared to the expected ppm-level experimental precision, enabling enhanced sensitivity or, in the absence of a signal, competitive limits to be set. 

\section{Projected bounds for JLab kinematics}\label{sec5}

We also made an assessment of the parameter space that can be probed at JLab, where the scan over the range of $m_X$ can be performed with the initial state radiation technique. The resulting limits, obtained by requiring the BSM contribution to remain within the projected experimental uncertainty of 1 ppm, are shown in Fig.~\ref{fig:ExLimit}, together with existing bounds from $\left(g-2\right)_e$ measurements \cite{Fan:2022eto,Pustyntsev:2025nwm} and visible decay constraints on dark photons from A1@MAMI \cite{Merkel:2014avp}, BaBar \cite{BaBar:2014zli}, NA48/2 studies of $\pi^0$ decays \cite{NA482:2015wmo}, and the NA64 analyses \cite{NA64:2019auh}. We also include the projected sensitivity from the Belle II analysis \cite{Ferber:2015jzj}, as well as the LHCb Run 3 projection based on studies of dark photon radiation from charm mesons \cite{Ilten:2015hya}.

The NA64 bound for visible decaying mediators was also translated to the scalar case. Additionally, we incorporate projections for the Mu3e experiment based on the $\mu^+ \to e^+\bar{\nu}_\mu \nu_e\tilde{S}$ decay, assuming $2.5 \times 10^{15}$ and $5.5 \times 10^{16}$ produced $\mu^+$, respectively (labeled I and II in Fig. \ref{fig:ExLimit}), adapted from \cite{DiLuzio:2025ojt}. We further reproduced limits from the SINDRUM experiment \cite{SINDRUM:1989qan}, which searched for the $\pi^+ \to e^+\nu \tilde{S}$ process, alongside the projected sensitivity of PIONEER, which aims to improve the branching ratio limit on this decay to $\mathcal{O}\left(10^{-11}\right)$. In both aforementioned processes, the scalar $\tilde{S}$ is emitted from the final-state positron and subsequently decays as $\tilde{S}\to e^+e^-$. 

Our results indicate that up to an order of magnitude extension in reach, as compared to current bounds, can be achieved for the scalar and vector scenarios for masses around 100 MeV without additional model assumptions. The ultimate reach, however, will depend on the precision achieved at JLab. 

We also emphasize that while recent missing-energy searches at NA64~\cite{NA64:2023wbi} have placed severe constraints on the parameter space of vector mediators decaying invisibly into dark sector states, these bounds are inherently model-dependent. The NA64 invisible limit in Fig.~\ref{fig:ExLimit} (gray dot-dashed line) assumes a dark coupling $\alpha_D=0.1$ and dark fermion masses $m_\chi = m_{A}'/3$. These constraints do not apply to the minimal scenario considered in this work, where the mediator decays with a $100\%$ branching ratio into visible $e^+e^-$ pairs; however, they show the complementarity of the two search strategies.

\section{Conclusion}\label{sec6}

In conclusion, we have shown that the Bhabha beam normal spin asymmetry, which can be accessed  using a polarized positron beam at JLab, provides a uniquely sensitive probe of BSM mediators. The uniqueness is threefold: firstly the JLab energy region, which corresponds with a maximum c.m. energy of 106 MeV of the Bhabha process prevents any hadronic discontinuity to contribute. Secondly, the leading QED result has a zero in the angular distribution due to cancellations between different discontinuities, allowing for a kinematic point which is nearly background free.  
Thirdly, the enhanced sensitivity of the observable $B_n$ to a BSM contribution results from the quadratic dependence on the coupling, in contrast to BSM searches in unpolarized processes, for which the BSM signal goes with the fourth power of the small coupling constant. 

We have shown that up to an order of magnitude extension in reach can be obtained, in particular for the scalar and vector scenarios. Our estimates were done in a conservative way, and additional improvements in energy resolution or experimental precision could strengthen the final results.
While the analysis presented in this work focused on the most natural scenarios involving scalar, vector, and axial vector mediators, the method can be straightforwardly extended to a broader class of models, including tensor-coupled interactions~\cite{Kozhuharov:2022qvw}. 

In this work we assumed a minimal model with a $100 \%$ branching ratio for the messenger decay into electron-positron pairs. If invisible decays into dark sector states are considered possible, the proposed framework remains fully applicable. However, since a larger total width $\Gamma_X$ suppresses the visible signal, the resulting constraints must be reinterpreted as bounds on the effective coupling scaled by the branching ratio, rather than the electron coupling alone, provided the total width $\Gamma_X$ remains smaller than the experimental invariant mass resolution.

Finally, also exploring the potential for BSM searches in double-polarization observables of the Bhabha process may be of interest for a future study.  
 
\section*{Acknowledgments}
The authors would like to thank Dave Mack, Cornelis J.G. Mommers, Mikhail Gorchtein, and Vladimir Pascalutsa for useful discussions. This work was made possible by Institut Pascal at Universit\'e Paris-Saclay with the support of the program “Investissements d’avenir” ANR-11-IDEX-0003-01. 
Furthermore, this work was supported by the Deutsche Forschungsgemeinschaft (DFG, German Research Foundation), in part through the Research Unit [Photon-photon interactions in the Standard Model and beyond, Projektnummer 458854507 - FOR 5327], and in part through the Cluster of Excellence [Precision Physics, Fundamental Interactions, and Structure of Matter] (PRISMA$^+$ EXC 2118/1) within the German Excellence Strategy (Project ID 39083149).

\bibliography{bibliography}

\end{document}